\documentclass[runningheads]{llncs}

\usepackage{cite}
\usepackage{amsmath,amssymb,amsfonts}
\usepackage{graphicx}
\usepackage{textcomp}
\usepackage[dvipsnames]{xcolor}
\usepackage{hyperref}
\usepackage{graphicx} 
\usepackage{adjustbox}
\usepackage{epstopdf}
\usepackage{verbatim}        
\usepackage{array}
\usepackage{nomencl}
\usepackage{times}        
\usepackage{enumerate}
\usepackage{multirow}
\usepackage{wrapfig}
\usepackage{paralist}
\usepackage{subcaption}
\usepackage{listings}
\usepackage{url}
\usepackage[]{hyperref}
\usepackage{rotating}
\usepackage{wasysym}
\usepackage{flushend}
\usepackage{eqparbox}
\usepackage{stmaryrd}
\usepackage{amsfonts}
\usepackage{tabularx}
\usepackage{booktabs}
\usepackage{tikz, calc, tikz-qtree}
\usepackage[ruled,linesnumbered,vlined]{algorithm2e}

\usepackage[font=small,skip=8pt]{caption}

\def\BibTeX{{\rm B\kern-.05em{\sc i\kern-.025em b}\kern-.08em
    T\kern-.1667em\lower.7ex\hbox{E}\kern-.125emX}}

\hypersetup{
	pdftoolbar=true,        %
	pdfmenubar=true,        %
	pdffitwindow=false,     %
	pdfstartview={FitH},    %
	pdftitle={},    %
	pdfauthor={},     %
	pdfsubject={},   %
	pdfcreator={},   %
	pdfproducer={}, %
	pdfkeywords={}, %
	pdfnewwindow=true,      %
	colorlinks=true,       %
	linkcolor=Brown,          %
	citecolor=OliveGreen,        %
	filecolor=magenta,      %
	urlcolor=NavyBlue           %
}

\newcommand{\mypar}[1]{\smallskip\noindent\textbf{#1.}}

\def\traceName{\sigma}

\def\eventLog{L}

\newcommand{\aLog}[1]{
	[\traceName_1, \dots, \traceName_{#1}]}

\newcommand{\eqClass}[1]{\llbracket #1\rrbracket}

\begin{document}
\title{PMDG: Privacy for Multi-Perspective Process Mining through Data Generalization}
\titlerunning{Privacy for multi-perspective process mining through data generalization.}
\author{Ryan Hildebrant\inst{1} \and
Stephan A. Fahrenkrog-Petersen\inst{2,3} \and
Matthias Weidlich\inst{2}\and 
Shangping Ren\inst{4}}
\authorrunning{R. Hildebrant et al.}

\institute{University of California, Irvine, \email{rhildebr@uci.edu}\and
Humboldt-Universität zu Berlin, \email{\{fahrenks,weidlima\}@hu-berlin.de}\and
Weizenbaum Institute for the Networked Society \and
San Diego State University, \email{sren@sdsu.edu}}
\maketitle              %
\begin{abstract}
Anonymization of event logs facilitates process mining while 
protecting sensitive information of process stakeholders. Existing 
techniques, however, focus on the privatization of the control-flow. Other 
process perspectives, such as roles, resources, and objects are neglected or 
subject to randomization, which breaks the 
dependencies between the perspectives. Hence, existing techniques 
are not suited for advanced process 
mining tasks, e.g., social network mining or predictive monitoring. 
To address this gap, we propose PMDG, a framework to ensure privacy for 
multi-perspective process mining through data generalization. It provides 
group-based privacy guarantees for an event log, while preserving the 
characteristic dependencies between the control-flow and further process 
perspectives. 
Unlike existing privatization techniques that rely on
data suppression or noise insertion, PMDG adopts data generalization:
a technique where the activities and attribute values referenced in events are generalized
into more abstract ones, to obtain equivalence classes that are
sufficiently large from a privacy point of view.
We demonstrate empirically that PMDG outperforms state-of-the-art 
anonymization techniques, when 
mining handovers and predicting outcomes.

\end{abstract}

\section{Introduction}
Privacy-preserving process mining~\cite{elkoumy2021privacy} enables data-driven 
analysis of business processes, while protecting sensitive 
data about the individuals involved in process execution.  
To this end, existing techniques rely on the anonymization of an event log, 
which is commonly modeled as a set of traces, with each trace being a sequence of 
events that denote activity executions. 
In order to obtain a provable privacy guarantee, the traces of an 
event log are transformed. Here, existing techniques differ in terms of the 
adopted privacy guarantee 
and the properties preserved by these transformations. Anonymization of event 
logs may guarantee differential 
privacy~\cite{dwork2008differential} or rely on group-based notions, such as 
k-anonymity and its derivatives~\cite{10.1142/S0218488502001648}. Moreover,
the respective transformations may only suppress behavior in the log or 
potentially introduce new and noisy behavior in terms of unseen sequences of 
activity executions.

Most techniques for privacy-preserving process 
mining~\cite{FahrenkrogPetersen2019PRETSAEL,RAFIEI2021101908,mannhardt2019privacy}
focus on the construction of a process model from an event 
log~\cite{augusto2018automated}. As such, 
they target the control-flow perspective of the process, trying to ensure 
that the anonymized event log includes semantically correct sequences of 
activity executions from the process. However, event logs also contain 
information on other process perspectives, such as roles, resources, and case 
objects. Advanced process mining tasks exploit the relation between the 
control-flow and these additional perspectives, e.g., to extract 
hand-overs during 
process execution~\cite{zhao2014process} or to construct models to predict 
the outcome of running process 
instances~\cite{DBLP:journals/tkdd/TeinemaaDRM19}. 

\begin{figure}[t!]
  \centering
  \subfloat[]{\label{fig:log}
  	\scriptsize
  	\begin{tabular}{l l l l }
  		\toprule
  		Case & Activity  &  Location & Role \\
  		\midrule
  		07 & Register & Day Clinic  & Admin \\
  		07 & Vitals &  Day Clinic & GP \\
  		07 & Consultation & Day Clinic & GP \\
  		07 & CT Scan  & Hospital & CA	\\  
  			... & ... & ... & ... \\
  		08 & Register & Hospital & Admin  \\
  		08 & Consultation & Hospital & CA	  \\
  		08 & MRI Scan & Hospital & CA 	\\  
  		... & ... & ... &... \\
  		\midrule
  		\multicolumn{4}{l}{\emph{Noise trace added by privatization:}}\\ 
  		\midrule
  		55 & Register & Hospital & Admin \\
  		55 & MRI Scan & Day Clinic & GP	\\  
  		55 & Consultation & Day Clinic & Admin\\
  		... & ... & ... &... \\
  		\bottomrule
   	\end{tabular}
  }\hfill
  \subfloat[]{\label{fig:graph}
      \includegraphics[width=0.58\linewidth]{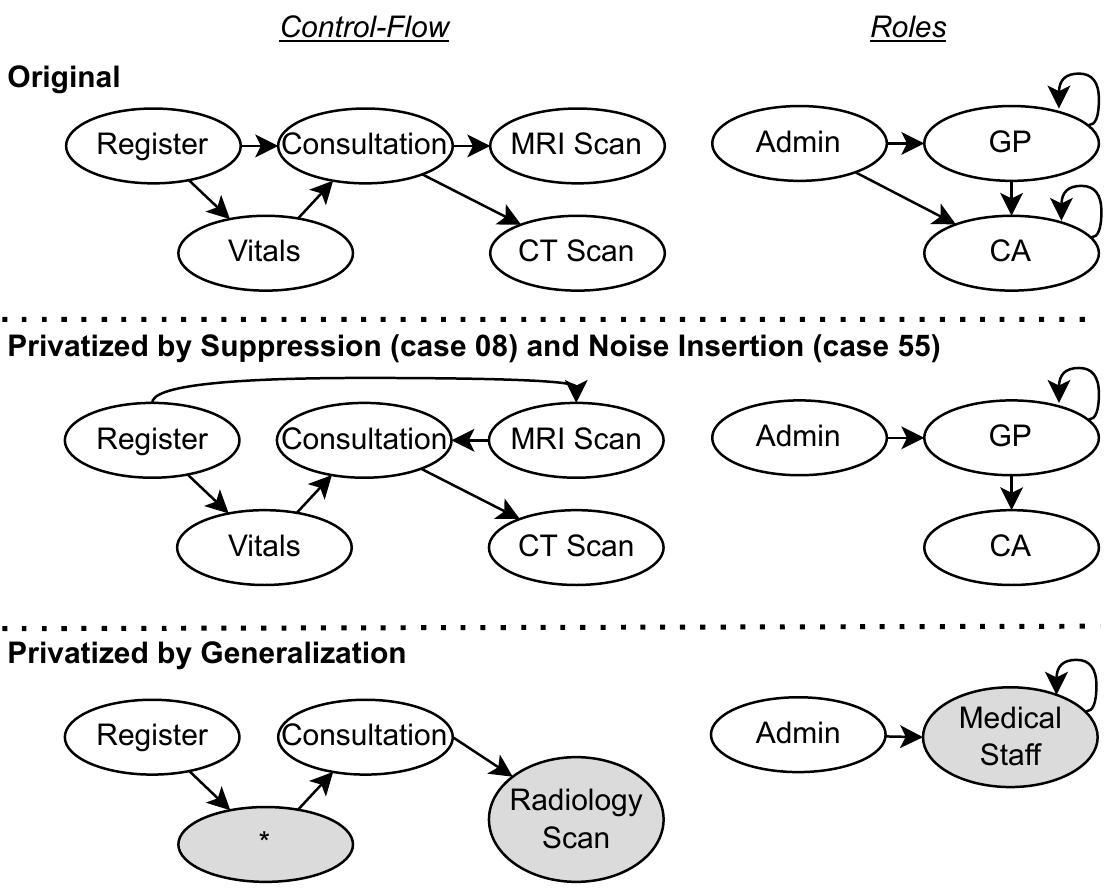}}
  \vspace{-0.6em}
  \caption{(a) Example log of a clinical pathway containing traces for case 
  07 and 08; (b) the control-flow and the role perspective, 
  when considering the original log, the event log privatized by suppression 
  (case 08 is suppressed) and noise insertion (case 55 is added), 
  and the event log privatized by generalization of activities and roles.}
  \label{fig:ex}
  \vspace{-1.8em}
\end{figure}

As of today, ensuring privacy beyond the control-flow creates a notable 
research gap. So far, data linked to events is either neglected, or assigned 
randomly once the control-flow has been 
anonymized~\cite{FahrenkrogPetersen20}. The latter tends to break any 
dependencies between the various perspectives, rendering the event 
logs unsuitable for multi-perspective process mining, as illustrated in 
\autoref{fig:ex}. Here, \autoref{fig:log} shows an example event log of a 
clinical pathway for two patients, cases 07 and 08. \autoref{fig:graph} (top), 
in turn, highlights the control-flow dependencies and the hand-overs between 
the involved roles. Existing techniques for privatization of the control-flow 
by suppression (of case 08) and noise insertion (of case 55), however, do not 
only disturb the control-flow, but also break the dependencies between process 
perspectives, as illustrated for the hand-overs between roles in 
\autoref{fig:graph} (middle).
This raises the question of \emph{how to preserve the characteristic 
dependencies between the process perspectives when 
anonymizing an event log}.

In this paper, we address this question with PMDG, a framework to ensure 
privacy for multi-perspective process mining through data generalization. 
It preserves the 
dependencies between process perspectives when constructing an event 
log that meets k-anonymity, a 
privacy guarantee often adopted in industry~\cite{kessler2019sap}. 
To this end, we adopt data generalization instead of data suppression and noise 
insertion. 
\autoref{fig:graph} (bottom) gives the intuition of this 
approach: here, sufficiently large equivalence classes of traces are derived by 
generalizing activities (\emph{CT Scan} and \emph{MRI Scan} become 
\emph{Radiology Scan}) and role information (\emph{GP} and 
\emph{CA} become \emph{Medical Staff}). While the generalization incurs some 
information loss, it arguably preserves general characteristics, such as the 
conduct of radiologic scans only after the consultation, as well as 
the handovers between administrative staff and medical personnel.

In sum, our contributions are the definition of PMDG as a first framework to 
enable privacy for multi-perspective process mining; and 
its instantiation with specific techniques for (i) the 
vectorization of traces to facilitate control-flow generalization; (ii) the 
selection of hierarchies to be used for the abstraction; and 
(iii) the application of the selected hierarchies to generalize the 
control-flow and the data assigned to events.

We demonstrate the effectiveness of PMDG for multi-perspective process mining 
in experiments with three public event logs. When mining handovers and 
predicting process outcomes, we observe that PMDG significantly outperforms 
state-of-the-art anonymization techniques in terms of maintaining 
characteristic hand-overs and classification accuracy, respectively. 
In the remainder, \autoref{sec:related_work} reviews related work on 
privacy-preserving process mining. 
\autoref{sec:background} then provides background information. The {PMDG} 
framework is outlined and instantiated in 
\autoref{sec:approach}, before we present evaluation experiments in 
\autoref{sec:poc}. We discuss our 
approach on a qualitative level in \autoref{sec:discussion}, before concluding 
in \autoref{sec:conclusion}. 

\section{Related Work}
\label{sec:related_work}
Privacy-preserving process mining has received much attention
recently~\cite{elkoumy2021privacy}.
Several approaches have been proposed to ensure $k$-anonymity and
other group-based privacy guarantees, e.g., by merging similar
traces~\cite{FahrenkrogPetersen2019PRETSAEL, batista2022privacy} or filtering
data~\cite{RAFIEI2021101908}. Due to their focus on the
control-flow, these methods are not suited for multi-perspective
process mining.

Instead of hiding sensitive data within groups
of traces, some approaches achieve differential privacy by inserting noise
into event
logs~\cite{mannhardt2019privacy,fahrenkog2021sacofa,elkoumy2021mine}. Here, the
privacy guarantees limits the effect one individual can have on the anonymized
data. Yet, the approaches filter behavior from the log or
introduce new and formerly unseen behavior.

The importance of the privatization of additional process perspectives has been
highlighted in~\cite{mining_roles}, which introduced a technique that is
tailored to one particular perspective, i.e. resource assignments.
In the general case, the
aforementioned approaches for control-flow anonymization may be combined with
an enrichment step, which either assigns values
randomly~\cite{FahrenkrogPetersen20} or unifies their distribution over an
event log~\cite{batista2021uniformization}. Either way, characteristic
dependencies between the process perspectives are compromised and the insertion
of new dependencies may lead to wrong conclusions in the analysis.

Another angle is followed in confidential process mining that aims to
protect an event log by encryption~\cite{rafiei2018supporting}. This
may include several process perspectives, but lacks any formal guarantee on
the privacy of individuals in the dataset.

Our approach relies on data generalization to privatize several process
perspectives. Generalization is a well-established method to privatize
relational, hierarchical, and simple sequence data, see~\cite{corpet1988multiple, wang2004bottom, 10.1145/1807167.1807248}.
Yet, PMDG is the first use of data generalization for event logs, i.e.
multi-variate sequences for which the dependencies between the various
dimensions shall be preserved.

\section{Background}
\label{sec:background}

Below, we summarize common notions for event logs and group-based privacy 
guarantees, as they are required for the definition of PMDG. 

\mypar{Event Logs}
Process mining is based on events, each representing the recorded execution of 
an {activity}, i.e. \emph{Register} or \textit{MRI Scan} in 
\autoref{fig:ex}. We denote the universes of activities and events by 
$\mathcal{A}$ and $\mathcal{E}$, respectively. The activity for which an event 
$e\in \mathcal{E}$ signals the execution is written as 
$e.a \in \mathcal{A}$. 
Events have a schema, defined by a set of attributes, $\mathcal{D} = 
\{D_1,\ldots,D_n 
\}$, and we denote the domain of values of attribute $D$ by  $\mathcal{V}_D$. 
For an event $e$, we write $e.D \in \mathcal{V}_D$ for the respective value 
of attribute $D$. In 
\autoref{fig:ex}, we have $\mathcal{D} = \{\mathit{Location}, \mathit{Role}\}$. 
Here, 
attribute $\mathit{Role}$ assumes the values $\mathit{Admin}$, 
$\mathit{CA}$, and $\mathit{GP}$, whereas the respective domain may also 
include further values. In particular, it can contain more abstract roles, such 
as $\mathit{Medical}\, \mathit{Staff}$ or $\mathit{Staff}$. 

Events that relate to the same, single execution of a process are grouped into 
a {trace}. Each trace $\traceName$ is a finite sequence of 
events $\langle e_1 , \dots, e_n \rangle\in \mathcal{E}^*$ of length 
$|\traceName|=n$. 
We  
use $\traceName.A$ to denote the {control-flow} of a trace $\traceName$, 
meaning the sequence of activities for which the execution is indicated by the 
events within the trace. For our running example, for instance, we have  
$\traceName.A = \langle \mathit{Register}, \mathit{Vitals}, 
\mathit{Consultation}, \mathit{CT\: Scan}\rangle$ for the trace of case 07. All 
traces with the same 
control-flow are said to be of the same trace {variant}, which is identified by 
one of the respective traces. That is, 
$\eqClass{\traceName}^A$ is the bag of 
traces that have the same control-flow as ${\traceName}$. 
A bag of traces is called an event log, $\eventLog = \aLog{n}$. It represents 
the input for many process mining algorithms.

\mypar{k-Anonymity}
A well-known way to protect the privacy of individuals is to hide them within a 
group, which is the aim of the $k$-anonymity privacy 
guarantee~\cite{10.1142/S0218488502001648}. 
The idea is that, in a dataset (an event log in our setting), one individual 
shall be indistinguishable from 
at least $k - 1$ other individuals. Therefore, the probability of identifying 
one individual, the so-called problem of \emph{identity disclosure}, can be 
bound to ${1}/{k}$. To achieve that $k$ individuals are indistinguishable, the 
{quasi-identifiers} need to be aligned. In general, quasi-identifiers are 
attributes that enable the identification of an individual, such as a postcode 
or birth date. 

In our setting, we 
assume that all attributes of all events and the control-flow of a trace can 
serve as quasi-identifiers. We therefore consider all traces that have the same 
control-flow and sequence of selected attribute values to be part of an {equivalence 
class}. The selected attributes are generated based on a defined perspective required for advanced process mining tasks. In line with the notation introduced for trace variants, we identify an 
equivalence class by one of its members, i.e. $\eqClass{\traceName \mid 
\mathcal{D}'}$ denotes an 
equivalence class that comprises all traces that have the same control-flow and 
attribute values for the specified attributes $\mathcal{D}'\subseteq 
\mathcal{D}$ as ${\traceName}$. 
Based thereon, we 
define $k$-anonymity as follows:
Let $\eventLog = \aLog{n}$ be an event log. Then, the log $L$ satisfies 
\emph{k-anonymity} with respect to a given perspective, 
if for every equivalence class $\eqClass{\traceName\mid \mathcal{D}'}$ in $L$ 
induced by a trace $\sigma \in L$, it holds that 
$\left|\eqClass{\traceName\mid \mathcal{D}'}\right| \geq k$. 

We note that the above definition of equivalence classes, and hence of 
k-anonymity, induces the strictest possible notion. It assumes the strongest 
adversary, under which any attribute and the control-flow may serve as 
quasi-identifiers in an identity disclosure attack. As such, it subsumes 
attacks in which an adversary possesses only a certain type of background 
knowledge~\cite{RAFIEI2021101908}, such as knowing which activities have 
been executed by an individual, but not the order of their execution. 
In order to avoid assumptions on the background knowledge of an adversary, 
we adopt the above model that represents the worst case scenario. 

\section{Generalization of Event Logs}
\label{sec:approach}
This section first
outlines the design principles for our work (\autoref{sec:design_principle}).
Then, we give an overview of our PMDG framework to address the identified
research gap (\autoref{sec:framework}). While some steps of it
rely on existing techniques, some aspects call for new techniques to ensure
high utility of the anonymized event log. Specifically, we introduce strategies
for trace vectorization (\autoref{sec:trace_vectorization}) and hierarchy
selection (\autoref{sec:hierarchy}).

\subsection{Design Principles of the Framework}
\label{sec:design_principle}
We developed the PMDG framework using the design science
methodology~\cite{peffers2007design}. The starting point for our problem
observation is that existing anonymization techniques for event logs mostly rely on noise insertion and aggregation, but do not incorporate any
generalization strategies. To address this research gap, we derived the
following
design objectives for the artifact: Given an event log with $m$
traces and a privacy parameter $k\leq m$, the artifact
(i) shall transform the event log to fulfill $k$-anonymity through
generalization; and
(ii) shall minimize the total amount of generalizations that are
applied to the log.
Within the remainder of this section, we introduce the PMDG framework to
realize these design objectives. The evaluation step of the artifact is
provided within a later section, while this paper denotes the final
{communication} step of the design science methodology.
\subsection{PMDG Framework}
\label{sec:framework}

\begin{figure}[t]
	\centering
	\includegraphics[width=0.7\linewidth]{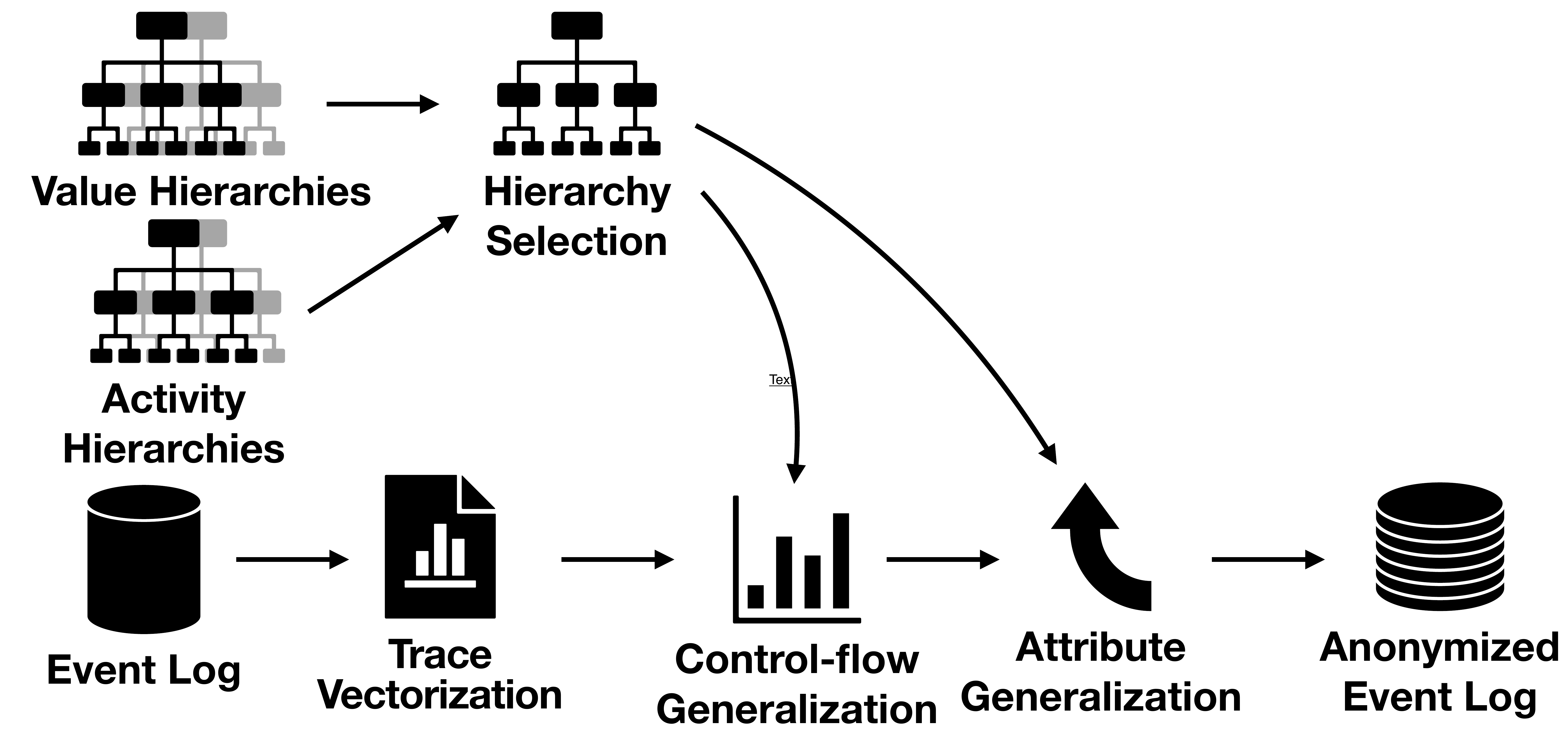}
	\caption{Overview of the PMDG framework.}
	\label{fig:PMDG}
	\vspace{-1.5em}
\end{figure}

As shown in \autoref{fig:PMDG}, our framework is applied to an event log that
contains information about
multiple process perspectives through the attribute values assigned to events
(see \autoref{sec:background}). In addition, it relies on generalization
hierarchies. That is, an \emph{activity hierarchy},
modeled as a function $\rho_A : \mathcal{A}
\rightarrow \mathcal{A'}\cup \{\star\}$, which maps an activity to a more abstract
activity or a wildcard $\star$. For an attribute $D$ representing an
an additional perspective, a \emph{value hierarchy} $\rho_{D}:
\mathcal{V}_D \rightarrow \mathcal{V}_D \cup \{\star\}$ maps an attribute value
to a more abstract value or a wildcard. Either way, for the control-flow or perspective,
the hierarchies are rooted in the wildcard $\star$, meaning that for any
activity $a\in \mathcal{A}$ and value $v\in \mathcal{V}_D$, it holds that
$\star\in \rho^*_A(a)$ and $\star\in \rho^*_D(v)$, where $\rho^*$ denotes
the transitive application of a generalization hierarchy $\rho$.
Moreover, for all process perspectives, multiple hierarchies may be available to
generalize the respective information.
Using the hierarchies, the PMDG framework transforms the event
log given as input, to one such that the resulting log guarantees $k$-anonymity.

Common strategies for data generalization are based on operations that change
individual values of the elements in a dataset~\cite{lefevre2005incognito}.
Hence, in order to enable comprehensive generalization, i.e. to achieve that
any two elements may end up in the same equivalence class, it is necessary to
ensure that all elements assume the same structure.
Transferred to
our setting, this requires all traces to be of equal lengths and all events to
have the same schema. While the latter requirement typically does not impose
any challenges in practice and is incorporated in our definition of the model
already, differences in the lengths of traces need to be handled. To this end,
our framework incorporates \emph{trace vectorization} as a first step, which we
explain in more detail in \autoref{sec:trace_vectorization}.

Next, the PMDG framework generalizes the control-flow before
considering further process perspectives. The reason being that
behavioral information serves as the basis for advanced process mining tasks
and shall always be protected by $k$-anonymity. Further perspectives enrich the
control-flow and may be subject to more fine-granular control of
the privacy guarantee, e.g., adopting
$t$-closeness~\cite{DBLP:conf/icde/LiLV07} instead of $k$-anonymity.

To obtain the privacy guarantee through generalization, however, multiple
hierarchies may be available, for the control-flow as well as for other process
perspectives. For instance, activities may be generalized according to the
artifact that is handled (e.g., all activities related to a \emph{CT Scan} are
generalized into a single activity) or by the type of the action that is
conducted (e.g., all activities that prescribe different drugs are combined
into a single activity). Similarly, roles may be generalized based on some
organizational structure (e.g., wards in a hospital) or some ability (e.g., the
specialization of a doctor). Each hierarchy will affect the utility of the
resulting event log differently. Hence, the PMDG framework includes a step to
guide the selection of one hierarchy for the control-flow generalization, and
one hierarchy for each attribute representing an additional process
perspective, as detailed in \autoref{sec:hierarchy}.

\begin{wrapfigure}{r}{0.54\textwidth}
	\vspace{-1.5em}
	\scriptsize
{\emph{Traces after trace
vectorization:}\vspace{.5em}}\newline
	\begin{tabular}{l l}
		\toprule
		\multicolumn{2}{c}{Case 07}\\
		 Activity  &  Role \\
		\midrule
		Register & Admin \\
		Vitals &  GP \\
		Consultation & GP \\
		CT Scan  & CA \\
		\bottomrule
	\end{tabular}\quad \quad \quad \quad \quad
	\begin{tabular}{l l}
		\toprule
		\multicolumn{2}{c}{Case 08}\\
		 Activity  &  Role \\
		\midrule
		Register & Admin  \\
		$\star$ &  $\star$ \\
		Consultation & CA	\\
		MRI Scan & CA 		\\
	\bottomrule
	\end{tabular}
\vspace{1em}\newline
{\emph{Traces after generalization:}\vspace{.5em}}\newline
	\begin{tabular}{l l}
		\toprule
		\multicolumn{2}{c}{Case 07}\\
		 Activity  &  Role \\
		\midrule
		Register & Admin \\
		$\star$ &  $\star$ \\
		Consultation & Medical Staff \\
		Radiology Scan  & CA \\
		\bottomrule
	\end{tabular}\quad
	\begin{tabular}{l l}
		\toprule
		\multicolumn{2}{c}{Case 08}\\
		 Activity  &  Role \\
		\midrule
		Register & Admin  \\
		$\star$ &  $\star$ \\
		Consultation & Medical Staff	\\
		Radiology Scan & CA 		\\
	\bottomrule
	\end{tabular}
	\caption{Generalization example.}
	\label{fig:ex2}
	\vspace{-8em}
\end{wrapfigure}

Once the generalization hierarchies have been selected, they are applied to the
event log using existing algorithms to achieve
$k$-anonymity~\cite{lefevre2005incognito,prasser2014arx}. In essence, these
algorithms adopt some step-wise generalization until the resulting equivalence
classes are sufficiently large. Specifically, in our context, given an
activity hierarchy $\rho_A$ and a value hierarchy $\rho_{D}$ per attribute
$D$, the result of the generalization step can be characterized as follows:
For each trace $\traceName=\langle e_1,\ldots,e_n \rangle $ of the original log
$L$, the resulting log $L'$ will contain a trace $\traceName'=\langle
e'_1,\ldots,e'_n \rangle$, such that for each event $e_j$, $1\leq j\leq n$, it
holds that the activity and each attribute value remains unchanged or has been
generalized, i.e. $e^\prime_j.a = e_j.a$ or $e_j.a=\rho^*_A(e_j.a)$ and
$e^\prime_j.D = e_j.D$ or $e_j.D=\rho^*_{D}(e_j.D)$.

For example, consider the traces for cases 07 and 08 from
\autoref{fig:ex} and focus solely
on the control-flow and the role perspective. Trace vectorization will
normalize the length of both traces by inserting an event with wildcard values
(discussed in \autoref{sec:trace_vectorization}), see \autoref{fig:ex2} (top).
With an activity hierarchy that
generalizes \emph{MRI Scan} and \emph{CT Scan} to \emph{Radiology Scan}, as
well as a
value hierarchy generalizing \emph{GP} and \emph{CA} to \emph{Medical Staff},
the traces can be generalized to fall into the same equivalence class, see
\autoref{fig:ex2} (bottom). As such, the resulting event log would satisfy
$2$-anonymity.

\subsection{Trace Vectorization}
\label{sec:trace_vectorization}

As explained above, comprehensive generalization requires that all elements in
a dataset assume the same structure.
Furthermore, we want to ensure that an anonymized event log
can be generated for every $k$ that is equal or smaller than the number of the
traces in the log. However, only traces with the same length can be merged into
the same
equivalence class. Therefore, we need to unify the length of traces in the
event log. To this end, we adopt trace vectorization, which is similar to the
idea of sequence encoding in predictive
process monitoring~\cite{leontjeva2016complex}.
Specifically, given an event
log $\eventLog = \aLog{n}$, trace vectorization
yields a log $\eventLog' = [\traceName'_1, \dots, \traceName'_{n}]$,
such that:
\begin{compactitem}
	\item All traces have the same length, i.e. for all $\sigma'_i,\sigma'_j$
	in $\eventLog'$, it holds $|\sigma'_i|=|\sigma'_j|$.
	\item For each trace $\sigma = \langle e_1,\ldots,e_m \rangle$ of the
	original log $L$, there is a corresponding trace $\sigma' = \langle
	e'_1,\ldots,e'_k \rangle $ in $L'$, so that the projection of $\sigma'$
	on the events $\{ e_1,\ldots,e_m \}$ yields the trace $\sigma$ and all
	events $e$ of $\sigma'$ that are not part of the projection are wildcard
	elements, i.e., it holds that $e.A = \star$ and $e.D = \star$ for all $D\in
	\mathcal{D}$.
\end{compactitem}
One naive approach for trace vectorization would be to extend all traces that
are shorter than the maximum length of traces in an event log with wildcard
events at the end. However, such an approach cannot be expected to preserve the
utility of the traces for process mining, especially considering the
control-flow perspective. For instance, for the trace of case 08 in
\autoref{fig:ex2}, adding the wildcard event at the end would have severe
consequences for the subsequent generalization: Instead of preserving the
information on the \emph{Consultation} and \emph{Radiology Scan} activities of
both traces, all except the first activity would be generalized to the root
element ($\star$).

In PMDG, therefore, we employ a strategy based on multi-sequence alignments
(MSA)~\cite{MSA}. In essence, MSA identifies how to insert gaps into sequences
of symbols, such that the same symbol is assigned to a certain index in all
sequences and the number of gap indices is minimal. In our setting, we adopt
MSA for the control-flow perspective, as it serves as the basis for process
mining tasks. That is, given an event log $\eventLog = \aLog{n}$, MSA is
applied to the set $\{\sigma_1.A, \ldots , \sigma_n.A \}$ to identify where
wildcard events shall be inserted.

\subsection{Hierarchy Selection}
\label{sec:hierarchy}
As detailed above, multiple hierarchies may be employed to generalize the
control-flow or the data representing additional process perspectives. Below,
we first elaborate on types of hierarchies and their origin as well as their
implications for the utility of the anonymized event log. We then present a
heuristic solution to guide the selection of generalization hierarchies as part
of PMDG.

\mypar{Types of hierarchies} In general, one can distinguish two types of
hierarchies:
\begin{compactenum}[(i)]
\item \emph{Syntactic hierarchies} are obtained by suppressing a part of an
activity label or an attribute value. Common examples for syntactic hierarchies
are numeric values (e.g., postcodes \emph{'12489'} and \emph{'12555'} are
generalized to \emph{'12---'}) or dates (\emph{'10/2022'} and
\emph{'12/2022'} are generalized to \emph{'--/2022'}). However, one may also
consider activities and generalize, for instance, \emph{CT Scan} and \emph{MRI
Scan} to \emph{Scan} by suppressing the first token of the label.
\item \emph{Semantic hierarchies} generalize the meaning of an activity or
attribute value. An example would be the generalization of an attribute
capturing a city (\emph{'Berlin'}) into a country
(\emph{'Germany'}), larger region (\emph{'EU'}), or continent
(\emph{'Europe'}). The creation of semantic hierarchies requires domain
knowledge and these hierarchies are usually either user-defined or extracted
from a knowledge base. For activities in traditional business processes, for
instance, the MIT process handbook~\cite{malone2003organizing} defines
generalization hierarchies of activities.
\end{compactenum}
The selection of a hierarchy will impact the utility of the resulting event
log, even when considering only a single type of hierarchy. Taking up the
example of syntactic generalizations of dates, \emph{'11.2022'} and
\emph{'12.2022'} may be generalized not only to \emph{'--/2022'}, but also to
\emph{'11/---'} and \emph{'12/---'}, respectively, depending on which parts
to suppress.
Either generalization provides a different kind of information, which
influences the types of questions that can be answered with process mining for
the anonymized log.

\mypar{Selecting a hierarchy} Since the selection of certain hierarchies for
data generalization has significant implications, ideally, one would test all
available hierarchies for the control-flow and all attributes. Measuring the
quality of the resulting event logs based on a chosen utility measure, the best
combination of hierarchies can be determined. However, such a brute-force
approach is typically infeasible, due to the exponential number of hierarchy
combinations.
Therefore, in PMDG, we incorporate a heuristic strategy to guide the selection
of a generalization hierarchy independently for the control-flow and each
attribute. The heuristic is based on a notion of utility, for which we consider
the following instantiations:
\begin{compactitem}
\item The utility is given by the number of equivalence classes within an
anonymized event log. Here, the intuition is that a larger number of
equivalence classes in the anonymized log yields a better representation of the
variance in the original log.
\item The utility is inversely proportional to the differences in size of the
equivalence classes, i.e. the number of contained traces. Here, the motivation
is to preserve information on common behavior more
precisely than on uncommon behavior.
\end{compactitem}
Based on a specific notion of utility, the selection of a hierarchy per process
perspective may be guided by an estimated utility, as follows. Let
$\{\rho^1_D,\ldots, \rho^n_D\}$ be a set of hierarchies for an attribute $D$ (or,
analogously, for the activities). Then, for each hierarchy, we determine the
equivalence classes when considering \emph{only} the attribute $D$ and
\emph{one} level in the generalization hierarchy (i.e.,
$\rho^1_D,\ldots\rho^n_D$ are applied only once, not transitively). Let $u^i_1$
be the utility as determined for the equivalence classes obtained with
$\rho^i_D$, which, as mentioned above, may be defined by the number of classes.
Afterwards, the equivalence classes obtained with subsequent levels of the
hierarchies are assessed iteratively, yielding utility values $u^i_j$ for
hierarchy $\rho^i_D$ when incorporating it up to level~$j$. Per hierarchy
$\rho^i_D$, these utility values are summed up in a weighted manner, i.e., $u^i
= \sum_{j=1}^k w_j \cdot u^i_j$ with $k$ as the maximum depth of the hierarchy.
Using the weights $w_j$ enables us to give preference to different levels of
generalization, i.e., prioritizing the generalization from a city to a country,
over the one from a country to a continent. Finally, we select a hierarchy
$\rho^i_D$ for which the estimated utility $u^i$ is maximal over all
hierarchies $\{\rho^1_D,\ldots\rho^n_D\}$ for attribute $D$.

\section{Evaluation}
\label{sec:poc}
Within this section, we investigate how anonymizing event logs with PMDG impacts the utility of advanced process mining tasks.
Through an empirical evaluation, we show the feasibility and effectiveness of
PMDG. First, we will give an overview of the datasets used in our experiments
in \autoref{dataset}. Next, we outline our experimental setup, baseline, and
evaluation metrics in \autoref{sec:experimental_setup}. The
results of our experiments in \autoref{sec:results}.

\subsection{Datasets and Implementation}
\label{dataset}
For our experiments, we use three real-world event logs: BPIC
2013~\cite{DBLP:conf/bpm/2013bpic}, Road Traffic Fines~\cite{road_traffic}, and
the CoSeLoG~\cite{Buijs2014}. For each log, we excluded all variants that only
appear once. This ensures a reasonable setting for anonymization (where unique
traces would be problematic in any case). Certain experiments with advanced process mining
tasks required the existence of the same attribute in all events, in that case
we performed these experiments only with the BPIC
2013~\cite{DBLP:conf/bpm/2013bpic} and CoSeLoG~\cite{Buijs2014} event logs, since road traffic fines is missing such an attribute.

For all of our experiments, we
provide an open-source implementation on
GitHub.\footnote{\url{https://github.com/Ryanhilde/PMDG_Framework/}} The trace
generalization approach is implemented in Python.
For the generalizations of attributes, we used Java libraries from the ARX
project.\footnote{\url{https://arx.deidentifier.org}} For our experiments on
mining handovers, we relied on the organizational mining features of
PM4Py.\footnote{\url{https://pm4py.fit.fraunhofer.de}}
For our experiments on outcome prediction, we used
scikit-learn.\footnote{\url{https://scikit-learn.org}}

\subsection{Experimental Setup}
\label{sec:experimental_setup}
\mypar{Parameter settings}
In our experiments, we use different strengths for $k$-anonymity, with values of $k$ varying from $\{5, 10, 15, 20\}$. Furthermore, we use semantic hierarchies that we created manually. We also tested syntactic hierarchies, but these were always outperformed in terms of the amount of changes applied to the anonymized log. We selected our semantic hierarchies based on retained equivalence classes for the control-flow and minimum generalizations for the attributes.
In our experiments regarding predictive process monitoring,  we trained decision trees on 1,000 randomly generated 20/80 test-train splits.

\mypar{Baseline} As a baseline for some of our experiments, we used PRIPEL~\cite{FahrenkrogPetersen20}, a framework that transforms event logs to achieve  $\epsilon$-differential privacy~\cite{dwork2008differential}. The provided privacy guarantee is not directly comparable with $k$-anonymity, i.e. we cannot assume that a specific $k$-value will ensure the same amount of privacy as a setting in PRIPEL. However, PRIPEL is the best choice for comparison, as it is the only existing technique that is capable of handling all attribute values. In general, a lower value for $\epsilon$ corresponds to a stronger privacy guarantee. For our experiments, we consider two settings for PRIPEL in the two event logs, a weak privacy guarantee ($\epsilon=1.0$) and a strong one  ($\epsilon=0.1$). Furthermore, we set the pruning parameter of PRIPEL to $2$ in the weaker setting and to $20$ for the stronger one; and always set the maximum prefix-length to the mean of the trace variants. These two parameters are required by PRIPEL, due to the underlying control-flow anonymization technique~\cite{mannhardt2019privacy}.
Furthermore, we compare our trace vectorization technique based on MSA with a naive approach, that fills up all traces at the end with wildcards.

\mypar{Evaluation metrics}
We use the number of remaining variants as a metric, to measure the
control-flow preservation after the anonymization, which is shown in
\autoref{tab:control_flow_preservation}. To study the impact of the attribute anonymization on the utility of advanced
process mining tasks, we investigated
two advanced process mining techniques: the discovery of handovers~\cite{zhao2014process,Aalst16} and process outcome prediction~\cite{DBLP:journals/tkdd/TeinemaaDRM19}.

\setlength{\tabcolsep}{12pt} %
\begin{table}[t]
	\caption{Comparison of Control-flow Preservation}
	\label{tab:control_flow_preservation}
	\centering
	\footnotesize
	\begin{tabular}{llllll}
		\toprule
		\textbf{Log} &
	     Trace Vec. & $k=5$ & $k=10$ & $k=15$ & $k=20$\\

		\midrule
		CoSeLoG & MSA & \textbf{17} & \textbf{13} & \textbf{9} & \textbf{8} \\
		& Naive &  \textbf{17}  &  \textbf{13}  & 8 & \textbf{8}  \\

		\midrule
		BPIC 2013 &MSA& \textbf{82} & 31 & 23 & \textbf{23}\\
		& Naive& 65& \textbf{34} &\textbf{27} & 22 \\

		\midrule
		Traffic Fines  & MSA & \textbf{75} & \textbf{53} & \textbf{43} & \textbf{37}   \\
		& Naive & 12 & 12& 12& 12 \\
		\bottomrule
	\end{tabular}
	\vspace{-1.5em}
\end{table}
\setlength{\tabcolsep}{4pt} %

Through handover analysis, an analyst can investigate which attribute values directly follow each other within two events of the same trace. Often this analysis is performed on resource related attributes such as resource role or location. In order to quantify the results of our anonymization, we measure the preserved information of the generalized event logs compared to the original event log. We utilize an information preservation metric to capture the information loss due to generalization. Our metric is based on the intuition that generalizing a handover from its original relation (e.g. in the case of resource locations: Germany to China) to a generalized relation (e.g. Europe to China) still has some utility. Furthermore, this utility is higher than if the handover would have been generalized to an even higher level (Europe to Asia) or the highest level (Europe to World). We therefore define the preservation for generalized handovers $p$ as:
\begin{equation}
    p = \frac{[1 - \frac{\alpha(e'_{1}.D)}{\alpha(*)} + \frac{\alpha(e_1.D)}{|\alpha(*)}] +
    [1 - \frac{\alpha(e'_{2}.D)}{\alpha(*)} + \frac{\alpha(e_{2}.D)}{\alpha(*)}]}{2}
\end{equation}
The assumption here is that $\alpha$ is a function that returns the number of
potential attribute values that can be represented through a (generalized)
attribute value, i.e. the value \emph{'EU'} could represent $27$ countries in
an attribute $D$ that encodes countries. The special case $\alpha(*)$ returns the number of fine-granular values of an attribute $D$, i.e. all possible countries. The values $e_1$ and $e_1'$ represent the original and generalized values for the left-side of the handover, respectively, while $e_2$ and $e_2'$ represent the right-side, i.e. (China). Our metric only measures the loss of
information for existing handovers, since generalization cannot insert new
handover relations.

As a second analysis task, we consider process outcome prediction. Here, the utility of an event log is given by the classification results. We assess these results using the well-known classification metrics:  \textit{precision}, the fraction of positively labeled instances that are actually correct;
\textit{recall}, the ratio of actual positives that are correctly labeled;  and \textit{F1-score}, the harmonic mean of precision and recall.

\subsection{Results}
\label{sec:results}
\mypar{Control-flow Preservation}
In ~\autoref{tab:control_flow_preservation}, we show that the MSA based trace vectorization usually outperforms the naive approach. We can see, it can provide significant benefits based on the traffic fines event logs. In cases where the naive approach is better, the benefit is comparatively small. Overall, we can observe that higher $k$ lead to a higher loss in control-flow variance.

\begin{figure}[h!]
	\vspace{-1em}
	\centering
	\includegraphics[width=\linewidth]{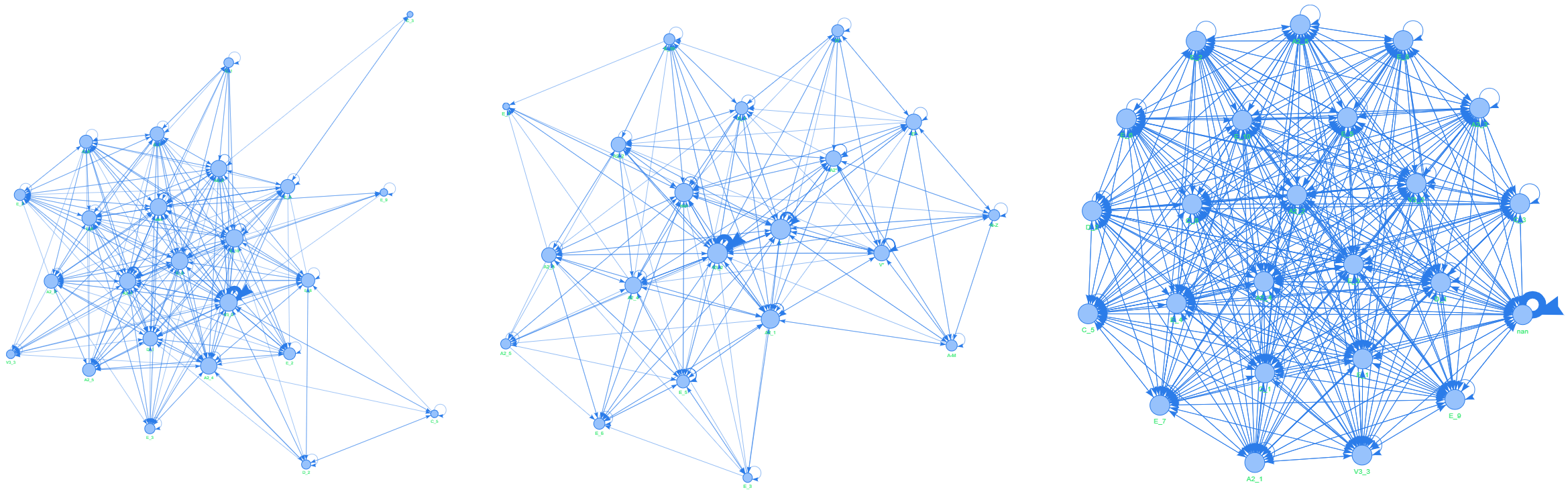}
	\caption{(Left) Original Event Log, (Middle) $k = 5$, and (Right) PRIPEL
	weak setting handover graphs for the org:role attribute.}
	\label{fig:orig_handover}
	\vspace{-1.5em}
\end{figure}

\mypar{Handovers}
In \autoref{fig:orig_handover}, we visualize the handovers created from the anonymized BPIC 2013 log based on the attribute \emph{org:role} and $k = 5$. Such an analysis would allow an organization to understand which kind of resource roles usually interact with each other. We can clearly see that the anonymization through PMDG produces a smaller handover graph (middle graph) that contains less information as compared to the original handover graph (left graph). However, more detailed insights can be derived from the results for the information preservation metric as shown in \autoref{tab:handover_results}. Here, we notice that a lot of handover information has actually been preserved. This highlights that the loss illustrated in \autoref{fig:orig_handover} is mostly due to the substitution of low-granularity handover relations with handover relations that are on a higher level of generalization.

In contrast,  the right graph in \autoref{fig:orig_handover} illustrates the
results obtained with PRIPEL using the weak configuration. Here, virtually all
attribute values are connected. While this, trivially, preserves all existing
handovers, it also introduces a large amount of false handovers.
Arguably, this is a major loss of information. However, this result is expected
for an anonymization technique that is based on noise insertion and that adopts
randomization for the attribute values assigned to events.

Let us illustrate the differences between the two
anonymization strategies with an exemplary analysis
question. That is, Volvo IT, the company from which the BPIC 2013 log
was obtained, was interested in understanding ping-pong behavior, i.e.,
cycles of handovers~\cite{DBLP:conf/bpm/2013bpic}.
Approaching this question based on the attribute \emph{org:role}, the original
log reveals handovers between roles E10 and
V3\_2, while there are no connections for the pairs of roles \{E9, V3\_2\},
\{A2\_1, C\_1\}, and \{A2\_2, C\_1\}. With PMDG, the roles E9 and E10 are
generalized into a single role E*, which is connected to role V3\_2. While this
hides the fact that E9 was not connected to V3\_2, it still suggests to
assess the handovers of the set of E roles with V3\_2. At the same
time, the graph with PMDG does not include the incorrect handovers for A2 and C\_1, so that
these roles are not considered in the analysis of ping-pong behavior. The
noisy result obtained with PRIPEL, in turn, is not suitable for this analysis,
as it suggests that all roles are involved in cyclic handovers.

\begin{table}[t]
	\caption{Precision of generalized handovers}
	\label{tab:handover_results}
	\centering
	\footnotesize
	\begin{tabular}{lllll}
		\toprule
		\textbf{Log and Attribute} &
	     $k=5$ & $k=10$ & $k=15$ & $k=20$\\

		\midrule
		BPIC 2013 "org:role" &  85.2 & 80.0 & 79.3 & 79.4 \\

		\midrule
		BPIC 2013 "organization involved" & 100 & 100 & 100 & 100 \\

		\midrule
		BPIC 2013 "resource country"  & 89.7 & 89.7 & 89.7 & 90.2   \\

		\midrule
		BPIC 2013 "organization country"  & 89.2 & 88.1 & 88.1 & 87.6   \\

		\midrule
		CoSeLoG "org:resource" & 73.1 & 75.4 & 75.3 & 75.3   \\
		\bottomrule
	\end{tabular}
	\vspace{-2em}
\end{table}

\begin{figure}
    \centering
    \includegraphics[height=\textheight]{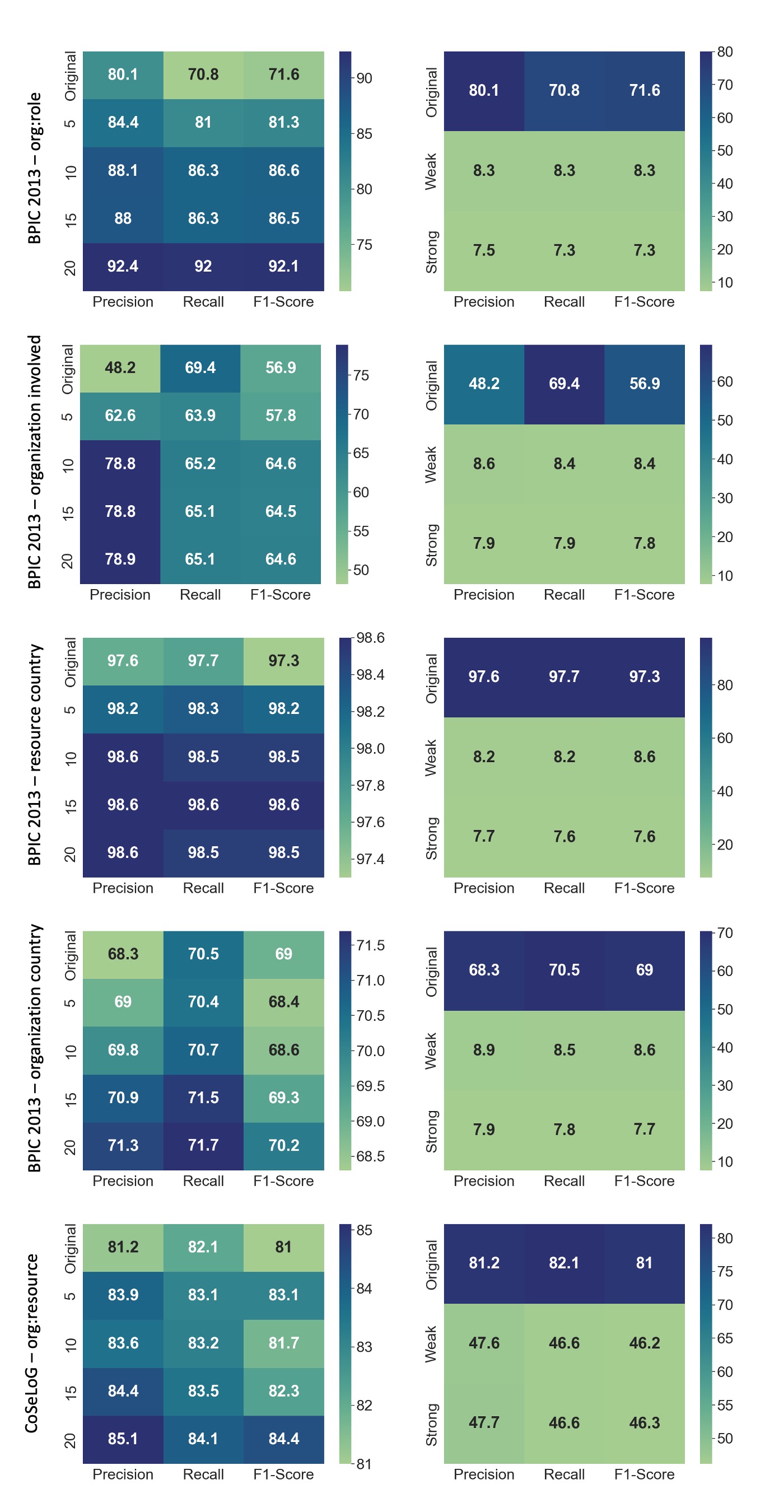}
    \caption{Decision Tree Results for the \textbf{PMDG} (left) and
    \textbf{PRIPEL} (right) approaches.}
	\label{fig:decision_tree}
	\vspace{-2.5em}
\end{figure}
\mypar{Process outcome prediction}
Next, we consider the common task of process outcome prediction. Here, we look
at the prediction of the ending activities  using a decision tree classifier.
In \autoref{fig:decision_tree}, we show the results from the classification
experiment.
The left heat maps show the results with different values for the privacy
guarantee $k$. We observe that for both the BPIC 2013 log
and the CoSeLoG log, higher privacy guarantees lead to better prediction
metrics. This behavior is expected, since a more general log contains less
control-flow variance, so that prediction becomes easier.

On the other hand, the classifiers trained based on event logs retrieved from PRIPEL provide classification results that have extremely low precision and recall. The results can be seen by observing the right heat maps. The noise inserted into the anonymized logs from PRIPEL clearly has a strong negative impact on the classification results and, hence, renders the anonymized event logs useless for outcome prediction.

\section{Discussion}
\label{sec:discussion}
Below, we discuss several aspects of our approach, which will help to understand opportunities and limitations that need to be considered when applying PMDG.

\mypar{Limitations of generalization}
The main drawback of generalization is the loss of detail
within the anonymized data. As a consequence, the data may no longer be useful
for certain analyses. For instance, if individual resources are generalized
towards their department, it is no longer possible to check whether the two-man
rule was followed. As another example, consider the medical domain, where the
dosage of a medication may be generalized. Questions related to the daily
dosage limit may become impossible to answer under strong generalization.

\mypar{The choice of generalization hierarchies}
The success of applying PMDG highly depends on the generalization hierarchies that are available. Semantic hierarchies require manual work for their creation, a factor that can limit their availability. Also, for certain attributes, it might not be obvious how to generalize them. A prominent example are activities that often lack an unambiguous generalization hierarchy. Without the knowledge of a domain expert, it is not clear  how to assess to which extent a generalization maintains process-specific information. Furthermore, PMDG makes no guarantee that the abstracted results will be useful in all situations. The usefulness of the results is dependent on the quality of the generalization hierarchy provided and the level of abstraction necessary to provide the privacy guarantee.

\mypar{Different levels of abstraction}
Based on the generalization technique used, an ano\-ny\-mized log might contain
attribute values with differing levels of abstraction, i.e. an attribute
encoding a region might contain values that represent a country or a continent.
Mixing these different levels of abstraction can be challenging in the
analysis, since most techniques do not offer built-in solutions to deal with
such heterogeneous abstraction levels. Therefore, event logs that are
anonymized with PMDG might require some post-processing before they can be
utilized in common process mining solutions.

\mypar{Risk of complete suppression}
If an event log only consists of variants with a small number of traces that
differ a lot in their attribute values, it is possible that these attribute
values are essentially suppressed, i.e., generalizing a region from a value representing a city to the value
\emph{'World'}. In such a case, all potential benefits of generalization are
lost. This problem can be addressed by providing hierarchies with a large
number of generalization levels, so that the attribute values can converge
to a level that still offers some utility.
However, a large number of generalization levels may lead to an event logs with a lot of
variance in its attribute values.

\mypar{Curse of dimensionality}
A well-known issue for achieving $k$-anonymity is the curse of
dimensionality~\cite{aggarwal2005k}, meaning that an increase in attributes or
events makes it harder to achieve the privacy guarantee. As we introduce
additional attributes assigned to case and events, the data is partitioned into
smaller equivalence classes. Consequently, an anonymized event log can be
expected to lose more utility. A
potential solution for this problem is the adoption of mixed privacy
guarantees~\cite{holohan2017k}. These techniques would allow for the use of
noise-based anonymization for some attributes and generalization for others,
while this choice is taken based on the requirements imposed in a specific
analysis setting.

\section{Conclusion}
\label{sec:conclusion}
Within this work, we introduced PMDG, an anonymization framework that transforms events logs, so that they are protected by $k$-anonymity. The novelty our approach comes from (i) its ability to preserve the dependencies between different process perspectives as recorded in an event log, i.e. the control-flow and the attribute values assigned to events; and (ii) the the utilization of data generalization techniques as a means to achieve a privacy guarantee.
In experiments with real-world event logs, we showed that PMDG outperforms the 
state of the art in terms of utility preservation for advanced process mining 
techniques. 
In future work, we intend to study 
how to support the construction and application of generalization hierarchies 
to optimize the utility of anonymized logs.
\vspace{-2.5em}
\section*{Acknowledgments}
\vspace{-1em}
 This work was supported by the NSF, grant number 1952225 and the German Federal Ministry of Education and Research (BMBF), grant number 16DII133.
 \vspace{-1em}
\typeout{}

\bibliographystyle{unsrt}

\end{document}